\documentclass[12pt]{article}
\usepackage{latexsym}
\usepackage{amssymb}
\usepackage[english]{babel}
\usepackage{amsmath}
\title{Quantum field description of the finite-width effects}
\author{V.I. Kuksa}
\date{Institute of Physics\\ Southern Federal University (former Rostov State University)\\
pr. Stachki 194, Rostov-on-Don, 344090 Russia, E-mail address:
kuksa@list.ru}
\begin{document}

\maketitle
\begin{abstract}

The model of unstable particles with random mass is suggested to
describe the finite-width effects. The phenomenological
manifestation of mass smearing is discussed in the framework of
the model.

\end{abstract}

\pagenumbering{arabic} \setcounter{page}{1}

\section{Introduction}

Quantum field description of the unstable particles (UP) with a
large width runs into some problems, which are under considerable
discussions \cite{1}. These problems have both the conceptual and
technological status and arise due to UP lie somewhat outside the
traditional formulation of quantum field theory \cite{2}. We can
not treat the UP with large width as asymptotic state and include
it into the set of initial or final states. Moreover, perturbative
approach is unfit in the resonance neighborhood. These conceptual
problems are connected with methodological difficulties, such as
an ambiguity in definition of mass and width. Therefore, the new
quantum field approach \cite{2} (Bohm et al), phenomenological
models \cite{3} and effective theories of UP \cite{4} are actual
now.

The convolution method \cite{5} is convenient and clear
phenomenological way to evaluate the instability or finite-width
effects (FWE). This method describes FWE in the processes
$\Phi\rightarrow\phi_1\phi\rightarrow\phi_1\phi_2\phi_3 ...$,
where $\phi$ is the UP with large width. The intermediate unstable
state $\phi$ is simulated by the final state $\phi$ in the decay
$\Phi\rightarrow\phi_1\phi$ with invariant mass, described by
Breit-Wigner-like (Lorentzian) distribution function. The
phenomenological expression for a decay rate has convolution form
\cite {5}:
\begin{equation}\label{E:1}
 \Gamma(\Phi\rightarrow\phi_1\phi) = \int_{q^2_1}^{q^2_2} \Gamma(\Phi\rightarrow
 \phi_1\phi(q))\rho(q)dq^2\,,
\end{equation}
where $\rho(q)=M\Gamma_{\phi}(q)/\pi|P(q)|^2$. In Eq.(\ref{E:1})
$\rho(q)$ is probability density of invariant mass distribution,
$P(q)=q^2-M^2+iM\Gamma_{\phi}(q)$ \cite{5}(Altarelli et al),
$\Gamma(\Phi\rightarrow\phi_1\phi(q))$ and $\Gamma_{\phi}(q)$ are
partial width of $\Phi$ and total width of $\phi$ in the stable
particle approximation, when $m^2_{\phi}=q^2$. The formula for a
decay rate, which has a close analogy to the Eq.(\ref{E:1}), was
applied to the description of FWE in $B$ and $\Lambda$ decay
channels with $\rho(770)$ and $a_1 (1260)$ in the final states
\cite{6}. It was shown that the contribution of FWE to the decay
rates of these channels are large (20-30 \%) and the account of it
significantly improves a conformity of experimental data and
theoretical predictions. Analogous results were obtained in Ref.
\cite{3} for the dominant decay channels of $\Phi(1020)$,
$\rho(770)$ and $K^*(892)$. The decay rates of the near-threshold
decay channels $t\rightarrow WZb, cWW, cZZ$ were calculated with
help of convolution formula (CF) in Ref. \cite{5}. It was shown in
these works, that the FWE play a significant role in the
near-threshold processes.

The convolution formula (\ref{E:1}) was derived in Ref. \cite{7}
by direct calculation with help of the decay-chain method. In this
work the contribution of all decay-chain channels of UP is
described by function $\rho(q)=q\Gamma(q)/\pi |P(q)|^2$. The
essential elements of this derivation for vector and spinor UP are
the expressions $\eta_{mn}=-g_{mn}+q_m q_n/q^2$ and
$\hat{\eta}=\hat{q}+q$ for numerators of vector and spinor
propagators ($\hat{q}=q_i\gamma^i$). The convolution formula was
derived for the decay chain $t\rightarrow bW\rightarrow bf_if_j$
in the limit of massless fermions $f$ in Ref. \cite{5} (Galderon
and Lopez-Castro). Quantitative analysis of convolution and
decay-chain calculations of the $t\rightarrow WZb$ decay rate was
fulfilled in Ref. \cite{5} (Altarelli et al). The formula for a
decay rate, which is similar to (\ref{E:1}), was received in Ref.
\cite{3} for the case of the scalar UP within the framework of the
"random mass" model. The UP is described in this model by the
quantum field with a "smeared" (fuzzy) random mass in accordance
with the uncertainty principle for energy and lifetime of unstable
quantum system \cite{8}. The FWE is connected with this
fundamental principle, which gives the relation $\delta m*\tau
\approx1$, that is $\delta m \approx \Gamma$ in the rest frame of
reference ($\delta E=\delta m$, $c=\hbar =1$) \cite{3}. So, the
uncertainty principle leads to the interpretation of kinematic
value $q^2$ in Eq.(\ref{E:1}) as a random mass square. Thus, the
intermediate states of UP, which are traditionally defined as
virtual, in the neighborhood of $q^2=M^2$ are not differ from real
ones in accordance with the uncertainty principle. This
interpretation is connected with a smearing of mass shell and with
above mentioned definition of $\eta_{mn}$ and $\hat{\eta}$, which
are proportional to the polarization matrix for the vector and
spinor UP (see section 3). As it was noted in Ref. \cite{7}, this
proportionality leads to the factorization of the expression for
width in the decay-chain method, and, as consequence, to the CF
(\ref{E:1}). Thus, the suggested model is theoretical framework of
the convolution method, which takes into account the uncertainty
principle.

In this paper we consider the generalization of the model
\cite{3}, which includes vector and spinor fields (Section 2).
Within the framework of this generalized model the CF is derived
for UP of arbitrary type (Section 3). To determine the probability
density $\rho(m)$, which is an analogue of $\rho(q)$ in
Eq.(\ref{E:1}), we put a connection between the model and
effective theory of UP with modified propagators, used in Ref.
\cite{7} (Section 4). It was shown in the section 4, that this
connection leads to Lorentzian probability density $\rho(m)$ and
to the traditional description of UP in the intermediate state by
dressed propagator, as a special case of suggested approach. The
model is applicable to the decay processes of type
$\Phi\rightarrow \phi_1\phi(q)$, that is describes UP in a final
state, and leads to the convolution formula (\ref{E:1}) for UP of
arbitrary type. In the Section 5 we have considered some examples
of FWE manifestations in a various regions of the particle
physics. The contributions of FWE (or mass smearing) into the
decay rates of $\phi(1020)\rightarrow K\bar{K}$, $B^0\rightarrow
D^-\rho^+$,  $W\rightarrow f_1\bar{f}_2$ decays and into the
oscillations in the systems of neutral mesons $M^0-\bar{M}^0$ are
evaluated within the framework of the model.

\section{The model of unstable particles with a random mass}

The effect of mass smearing is described by the wave packet with
some weight function $\omega(\mu)$, where $\mu$ is random mass
parameter \cite{3}. The model field function, which simulates UP
in the initial, final or intermediate states, is represented by
the expression:
\begin{equation}\label{E:2}
 \Phi_{\alpha}(x)=\int \Phi_{\alpha}(x,\mu)\omega(\mu)d\mu\,.
\end{equation}
In Eq.(\ref{E:2}) $\Phi_{\alpha}(x,\mu)$ are the components of
field function, which are determined in the usual way when
$m^2=\mu$ is fixed (stable particle approximation). The limits of
integration will be defined in the sections 3 and 4.

The model Lagrangian, which determines "free" unstable field
$\Phi(x)$, has the convolution form:
\begin{equation}\label{E:3}
 L(\Phi(x))=\int L(\Phi(x,\mu))|\omega(\mu)|^2\,d\mu\,.
\end{equation}
In Eq.(\ref{E:3}) $L(\Phi(x,\mu))$ is standard Lagrangian, which
describes model "free" field $\Phi(x,\mu)$ in stable particle
approximation ($m^2=\mu$).

From Eq.(\ref{E:3}) and prescription
$\partial\Phi(x,\mu)/\partial\Phi(x,\mu^{'})=\delta(\mu-\mu^{'})$
it follows Klein-Gordon equation for the spectral component:
\begin{equation}\label{E:4}
 (\square-\mu)\Phi_{\alpha}(x,\mu)=0.
\end{equation}
As a result we have standard momentum representation of field
function for fixed mass parameter $\mu$:
\begin{equation}\label{E:5}
 \Phi_{\alpha}(x,\mu)=\frac{1}{(2\pi)^{3/2}}
 \int\Phi_{\alpha}(k,\mu)\delta(k^2-\mu)e^{ikx}dk.
\end{equation}
All standard definitions, relations and frequency expansion take
place for $\Phi_{\alpha}(k,\mu)$, but the relation
$k^0_{\mu}=\sqrt{\bar{k}^2+\mu}$ defines smeared (fuzzy)
mass-shell due to random $\mu$.

The expressions (\ref{E:2}) and (\ref{E:3}) define the model
"free" unstable field, which really is some effective field. This
field is formed by interaction of "bare" UP with decay products
and includes nonperturbative self-energy contribution in the
resonant region. Such an interaction leads to the spreading
(smearing) of mass, that is to the transition from
$\rho^{st}(\mu)=\delta(\mu-M^2)$ for the bare particles to some
smooth density function $\rho(\mu)=|\omega(\mu)|^2$ with mean
value $\bar{\mu}\approx M^2$ and $\sigma_{\mu}\approx \Gamma$. So,
the UP is characterized in the discussed model by the weight
function $\omega(\mu)$ or probability density $\rho(\mu)$ with
parameters $M$ and $\Gamma$ (or real and imaginary parts of pole).
A similar approach has been discussed by Matthews and Salam in
Ref. \cite{8}.

The commutative relations for model operators have an additional
$\delta$-function:
\begin{equation}\label{E:6}
 [\dot{\Phi}^{-}_{\alpha}(\bar{k},\mu),\,\Phi^{+}_{\beta}(\bar{q},\mu^{'})]_{\pm}
 =\delta(\mu-\mu^{'}) \delta(\bar{k}-\bar{q})\delta_{\alpha\beta},
\end{equation}
where subscripts $\pm$ correspond to the fermion and boson fields.
The presence of $\delta(\mu-\mu^{'})$ in Eq.(\ref{E:6}) means an
assumption - the acts of creations and annihilations of particles
with various $\mu$ (random mass square) don't interfere. So, the
parameter $\mu$ has the status of physically distinguishable value
as random $m^2$. This assumption directly follows from the
interpretation of $q^2$ in Eq. (\ref{E:1}) as random parameter
$\mu$. By integrating both side of Eq.(\ref{E:6}) with weights
$\omega^{*}(\mu)\omega(\mu^{'})$ one can get standard commutative
relations
\begin{equation}\label{E:7}
 [\dot{\Phi}^-_{\alpha}(\bar{k}),\Phi^+_\beta(\bar{q})]_{\pm}=\delta(\bar{k}-\bar{q})
 \delta_{\alpha\beta}\,,
\end{equation}
where $\Phi^{\pm}_{\alpha}(\bar{k})$ is full operator field
function in momentum representation:
\begin{equation}\label{E:8}
 \Phi^{\pm}_{\alpha}(\bar{k})=\int\Phi^{\pm}_{\alpha}(\bar{k},\mu)\omega(\mu)d\mu\,.
\end{equation}
It should be noted that Eq.(\ref{E:7}) follows from Eq.(\ref{E:6})
when $\int|\omega(\mu)|^2d\mu=1$.

The expressions (\ref{E:2}) and (\ref{E:6}) are the principal
elements of the discussed model. The weight function $\omega(\mu)$
in Eq.(\ref{E:2}) (or $\rho(\mu)$) is full characteristic of UP
and the relations (\ref{E:6}) define the structure of the model
amplitude and of the transition probability (section 3). The
probability density $\rho(\mu)$ will be defined in the fourth
section by matching the model propagator to renormalized one.

With help of traditional method one can get from Eqs.(\ref{E:2}),
(\ref{E:4}) and (\ref{E:6}) the expression for the unstable scalar
Green function \cite{3}:
\begin{equation}\label{E:9}
 \langle 0|T(\phi(x),\phi(y))|0\rangle=D(x-y)=\int
 D(x-y,\mu)\rho(\mu)d\mu\,.
\end{equation}
In Eq.(\ref{E:9}) $D(x,\mu)$ is standard scalar Green function
with $m^2=\mu$, which describes UP in an intermediate state:
\begin{equation}\label{E:10}
 D(x,\mu)=\frac{i}{(2\pi)^4}\int\frac{e^{-ikx}}{k^2-\mu+i\epsilon}dk\,.
\end{equation}
The right side of the Eq.(\ref{E:9}) is Lehmann-like spectral (on
$\mu$) representation of the scalar Green function, which
describes the propagation of scalar UP. Taking into account the
connection between scalar and vector Green functions, we can get
the Green function of the vector unstable field:
\begin{equation}\label{E:11}
 D_{mn}(x,\mu)=-(g_{mn}+\frac{1}{\mu}\frac{\partial^2}{\partial
 x^n\partial x^m})D(x,\mu)=\frac{-i}{(2\pi)^4}\int\frac{g_{mn}-k_m
 k_n/\mu}{k^2-\mu+i\epsilon}e^{-ikx}dk\,.
\end{equation}
Analogously Green function of the spinor unstable field:
\begin{equation}\label{E:12}
 \Hat{D}(x,\mu)=(i\hat{\partial}+\sqrt{\mu})D(x,\mu)=\frac{i}{(2\pi)^4}
 \int\frac{\hat{k}+\sqrt{\mu}}{k^2-\mu+i\epsilon}e^{-ikx}dk\,,
\end{equation}
where $\hat{k}=k_i\gamma^i$. These Green functions in momentum
representation have a convolution structure:
\begin{equation}\label{E:13}
 D_{mn}(k)=\int D_{mn}(k,\mu)\rho(\mu)d\mu\,,\,\,\,\,\,
 \Hat{D}(k)=\int \Hat{D}(k,\mu)\rho(\mu)d\mu\,.
\end{equation}

\section{The model amplitude and the convolution formula for a decay
rate}

In this section we consider the model amplitude for the simplest
processes with UP in a final state and get the CF (\ref{E:1}) as
direct consequence of the model. The expression for a scalar
operator field \cite{3}:
\begin{equation}\label{E:14}
 \phi^{\pm}(x)=\frac{1}{(2\pi)^{3/2}}\int\omega(\mu)d\mu\int\frac{a^{\pm}(\bar{q},\mu)}
 {\sqrt{2q^0_{\mu}}}e^{\pm iqx}d\bar{q}\,,
\end{equation}
where $q^0_{\mu}=\sqrt{\bar{q}^2+\mu}$ and $a^{\pm}(\bar{q},\mu)$
are creation or annihilation operators of UP with momentum $q$ and
mass square $m^2=\mu$. Taking into account Eq.(\ref{E:6}) we can
get:
\begin{equation}\label{E:15}
 [\dot{a}^{-}(\bar{k},\mu),\phi^{+}(x)]_{-};\, [\phi^{-}(x), \dot{a}^{+}(\bar{k},\mu)]_{-}
 =\frac{\omega(\mu)}{(2\pi)^{3/2}\sqrt{2k^0_{\mu}}}e^{\pm ikx}\,,
 \,\,\,k^0_{\mu}=\sqrt{\bar{k}^2+\mu}\,.
\end{equation}
The expressions (\ref{E:15}) differ from standard ones by the
factor $\omega(\mu)$ only. From this result it follows that, if
$\dot{a}^{+}(k,\mu)|0\rangle$ and $\langle0|\dot{a}^{-}(k,\mu)$
define UP with a mass $m^2=\mu$ and a momentum $k$ in the initial
and final states, then the amplitude for the decay of type
$\Phi\rightarrow\phi\phi_1$ has the form:
\begin{equation}\label{E:16}
 A(k,\mu)=\omega(\mu)A^{st}(k,\mu)\,,
\end{equation}
where $A^{st}(k,\mu)$ is amplitude in a stable particle
approximation when $m^2=\mu$. This amplitude is calculated in a
standard way and can include high corrections. Moreover, it can be
effective amplitude for the processes with hadron participation
\cite{3,5}.

To define the transition probability of the process
$\Phi\rightarrow\phi\phi_1$, where $\phi$ is UP with a large
width, we should take into account the status of parameter $\mu$
as physically distinguishable value, which follows from
Eq.(\ref{E:6}). Thus, the amplitude at different $\mu$ don't
interfere and we have the convolution structure of differential
(on $k$) probability:
\begin{equation}\label{E:17}
 d\Gamma(k)=\int d\Gamma^{st}(k,\mu)|\omega(\mu)|^2d\mu\,.
\end{equation}
In Eq.(\ref{E:17}) the differential probability
$d\Gamma^{st}(k,\mu)$ is defined in the standard way (stable
particle approximation):
\begin{equation}\label{E:18}
 d\Gamma^{st}(k,\mu)=\frac{1}{2\pi}\delta(k_{\Phi}-k_{\phi}-k_1)|A^{st}
 (k,\mu)|^2d\bar{k}_{\phi}d\bar{k}_1\,,
\end{equation}
where $k=(k_{\Phi},k_{\phi},k_1)$ denotes the momenta of
particles. From Eqs.(\ref{E:17}) and (\ref{E:18}) it directly
follows the known convolution formula for a decay rate
\begin{equation}\label{E:19}
 \Gamma(m_{\Phi},m_1)=\int_{\mu_0}^{\mu_m}\Gamma^{st}(m_{\Phi},m_1;\mu)\rho(\mu)d\mu\,,
\end{equation}
where $\rho(\mu)=|\omega(\mu)|^2$ and $\mu_0,\mu_m$ are defined in
Refs. \cite{5,7} as threshold and maximal invariant mass square of
unstable $\phi$.

An account of high corrections in the amplitude (\ref{E:16}) keeps
the convolution form of Eq.(\ref{E:19}). This form can be
destroyed by accounting of the interaction between the products of
UP ($\phi$) decay and initial $\Phi$ or final $\phi_1$ states. The
calculation in this case can be fulfilled in a standard way, but
UP in the intermediate state is described by the model propagator.
However, a calculation within the framework of perturbative theory
(PT) can not be applicable to the UP with large width, that is to
the short-living particle. In any case, the applicability of the
PT, the model approach or convolution method to the discussed
decays should be justified by experiment. The correspondence of CM
to the experimental data was demonstrated for some processes
\cite{3,5,6,7}, but this problem needs in more detailed
investigation. In this connection we should note the analysis of
higher-order corrections for processes with UP \cite{4}. The
separation between factorizable and non-factorizable corrections
make it possible to build the effective theory of UP \cite{4}.

When there are two UP with large widths in a final state
$\Phi\rightarrow\phi_1\phi_2$, then in analogy with the previous
case one can get double convolution formula:
\begin{equation}\label{E:20}
 \Gamma(m_{\Phi})=\int\int\Gamma^{st}(m_{\Phi};\mu_1,\mu_2)\rho_1(\mu_1)\rho_2(\mu_2)d\mu_1
 d\mu_2\,.
\end{equation}
The derivation of CF for the cases when there is vector or spinor
UP in a final state can be done in analogy with the case of scalar
UP. However, in Eqs.(\ref{E:14}), (\ref{E:15}) and (\ref{E:16}) we
have a polarization vector $e_m(q)$ or spinor
$u^{\nu,\pm}_{\alpha}(q)$, where q is on fuzzy mass-shell. As a
result we get polarization matrix with $m^2=\mu$. For the vector
UP in a final state:
\begin{equation}\label{E:21}
 \sum_{e} e_m(q)e^{*}_n(q)=-g_{mn}+q_mq_n/\mu\,.
\end{equation}
For the spinor UP in a final state:
\begin{equation}\label{E:22}
 \sum_{\nu} u^{\nu,\pm}_{\alpha}(q)\bar{u}^{\nu,\mp}_{\beta}(q)=\frac{1}{2q^0_{\mu}}
 (\hat{q}\mp\sqrt{\mu})_{\alpha\beta}\,.
\end{equation}
In Eqs.(\ref{E:21}) and (\ref{E:22}) sum run over polarization and
$q^0_{\mu}=\sqrt{\bar{q}^2+\mu}$.

The formulae (\ref{E:19}) and (\ref{E:20}) describe FWE in full
analogy with the phenomenological convolution method \cite{5} and
with some cases of the decay-chain method \cite{5,7}. Thus, we
consider the quantum field basis for CM, which takes into account
the fundamental uncertainty principle and is in a good agreement
with the experimental data on some decays.  To evaluate FWE for
the case, when UP is in an initial state, we must account the
process of UP generation. When UP is in an intermediate state,
then the description of FWE is equivalent to the traditional one,
but the model propagators are determined by Eqs.(\ref{E:9}) -
(\ref{E:13}).

\section{Determination of random mass distribution function}

The possibility of $\rho(\mu)$-determination directly follows from
the connection of the decay-chain method (DCM) and convolution
method \cite{7}. As was shawn in Ref. \cite{7}, this connection
leads to the convolution formula (\ref{E:1}), where in accordance
with uncertainty principle we interpret the value $q^2$ as random
mass square parameter $\mu$, which distribution is described by
the expression:
\begin{equation}\label{E:23}
 \rho(\mu)=\frac{1}{\pi}\frac{\sqrt{\mu}\,\Gamma(\mu)}{|P(\mu)|^2}\,.
\end{equation}
In Eq.(\ref{E:23}) $\Gamma(\mu)$ is $\mu$-dependent full width and
$P(\mu)^{-1}$ is propagator's denomenator. It should be noted,
that the convolution structure of Eq.(\ref{E:1}) and universal
structure of Eq.(\ref{E:23}) don't depend on the definition of
$P(\mu)$. In general $P(\mu)$ has a complex pole structure
$\mu-\mu_R$ and can be approximated by the Breit-Wigner
$\mu-M^2+iM\Gamma(\mu)$ \cite{5} or another phenomenological
formulae. The expression (\ref{E:23}) is very simple and
convenient in practical calculations of decay rate, where the
error of approximation is small.

Here we'll consider the definition of $\rho(\mu)$ from the
matching model propagators to standard dressed ones \cite{3}. This
consideration is rather methodological than practical and
demonstrates the connection between model and traditional
descriptions. Let us associate the model propagator of scalar
unstable field (\ref{E:9}) with standard one:
\begin{equation}\label{E:24}
 \int\frac{\rho(\mu)d\mu}{k^2-\mu+i\epsilon}\longleftrightarrow
 \frac{1}{k^2-m^2_0-\Pi(k^2)}\,,
\end{equation}
where $\Pi(k^2)$ is conventional self-energy of scalar field. With
help of an analytical continuation of the expressions (\ref{E:24})
on complex plane $k^2\rightarrow k^2\pm i\epsilon$ and
prescription \cite{9}:
\begin{equation}\label{E:25}
 \Pi(k^2\pm i\epsilon)=Re\Pi(k^2)\mp iIm\Pi(k^2)
\end{equation}
the conformity (\ref{E:24}) can be represented by the equality
\begin{equation}\label{E:26}
 \int_{0}^{\infty}\frac{\rho(\mu}{k^2-\mu\pm
 i\epsilon}d\mu=\frac{1}{k^2-m^2(k^2)\pm
 iIm\Pi(k^2)}\,,
\end{equation}
where $m^2(k^2)=m^2_0+Re\Pi(k^2)$. With account of round pole
rules and $d\mu=d(\mu\mp i\epsilon), \rho(\mu\mp
i\epsilon)=\rho(\mu)\mp O(i\epsilon)$ two Eqs.(\ref{E:26}) can be
combined into the equality ($\mu\pm i\epsilon\rightarrow z)$:
\begin{equation}\label{E:27}
 \oint\frac{\rho(z)}{z-k^2}dz=\frac{1}{k^2-m^2(k^2)-iIm\Pi(k^2)}-
 \frac{1}{k^2-m^2(k^2)+iIm\Pi(k^2)}\,.
\end{equation}
The left side of Eq.(\ref{E:27}) is Cauchy integral, which equal
to $2\pi i\rho(k^2)$ and after a change $k^2\rightarrow\mu$ in the
final expression for $\rho$ we have:
\begin{equation}\label{E:28}
 \rho(\mu)=\frac{1}{\pi}\,\frac{Im\Pi(\mu)}{[\mu-m^2(\mu)]^2+[Im\Pi(\mu)]^2}\,.
\end{equation}
The expression (\ref{E:28}) for $\rho(k^2)$ in Breit-Wigner
approximation is usually exploited within the framework of
convolution method. From Eq.(\ref{E:28}) and definition
$\rho(\mu)=|\omega(\mu)|^2$ it follows:
\begin{equation}\label{E:29}
 \omega(\mu)=\frac{1}{\sqrt{\pi}}\,\frac{\sqrt{Im\Pi(\mu)}}{\mu-m^2(\mu)\pm
 iIm\Pi(\mu)}\,.
\end{equation}
The ambiguity of sign in (\ref{E:29}) is not essential because the
expression $|\omega(\mu)|^2$ only enters into the physical values.
In the parametrization $Im\Pi(\mu)=\sqrt{\mu}\,\Gamma(\mu)$ we
have relativistic Breit-Wigner $\omega(\mu)$ and Lorentzian
$\rho(\mu)$, which coincides with the expression (\ref{E:23}) for
renormalized  $P(q^2)$. Inserting the expression (\ref{E:28}) into
the left side of Eq.(\ref{E:24}) one can check with help of Cauchy
method the self-consistency of Eqs.(\ref{E:24}) and (\ref{E:28}).

Thus, we have put the correspondence between the model \cite{2} -
\cite{6} and some effective theory of UP with renormalized
propagator of scalar UP. To establish such a correspondence for
the vector UP we insert $\rho(\mu)$ into the model propagator
(\ref{E:13}) with $D_{mn}(k,\mu)$, defined by (\ref{E:11}) for
vector unstable field:
\begin{align}\label{E:30}
 \int_{0}^{\infty}\frac{-g_{mn}+k_m k_n
 /\mu}{k^2-\mu+i\epsilon}\,\frac{1}{\pi}\,\frac{Im\Pi(\mu)}{[\mu-m^2(\mu)]^2+
 [Im\Pi(\mu)]^2}\,d\mu=\\ \notag
 \frac{1}{2i\pi}\int_{0}^{\infty}\frac{-g_{mn}+k_m k_n
 /\mu}{k^2-\mu+i\epsilon}[\frac{1}{\mu-m^2(\mu)-i
 Im\Pi(\mu)}-\frac{1}{\mu-m^2(\mu)+iIm\Pi(\mu)}]d\mu\,.
\end{align}
With help of Eq.(\ref{E:25}) and above used method we can
represent the second part of Eq.(\ref{E:30}) in the
form($\mu\rightarrow z=\mu\pm i\epsilon$):
\begin{equation}\label{E:31}
 \frac{1}{2i\pi}\oint\frac{dz}{z-k^2}\,\,\frac{-g_{mn}+k_m k_n
 /z}{z-m^2(z)-iIm\Pi(z)}=\frac{-g_{mn}+k_m
 k_n/k^2}{k^2-m^2(k^2)-iIm\Pi(k^2)}\,.
\end{equation}
The right side of Eq.(\ref{E:31}) is similar to the expression for
propagator of vector UP, which leads to the convolution formula
(\ref{E:1}) in the decay-chain method \cite{7}. The numerator of
this effective propagator coincides with $\eta_{mn}(k)$, which was
used in \cite{7}. In Eqs.(\ref{E:30}) and (\ref{E:31}) the value
$\Pi(k^2)$ is defined for vector field as transverse part of
polarization matrix \cite{1}. The calculations of $\Pi(k^2)$ in
effective theory (unstable hadrons) or in gauge theory
(Z,W-bosons) can run into some difficulties. In the first case
loop calculation can be ambiguous and we should use traditional
Breit-Wigner approximation $m^2(\mu)\approx M^2$ and
$Im\Pi(\mu)\approx\mu\Gamma(\mu)$. To escape the gauge-dependence
in the second case we can use pole definitions of mass and width
\cite{1}.

The description of $\rho(\mu)$ by the universal function
(\ref{E:28}) for scalar and vector fields can be justified by the
general structure of parametrization for bosons:
\begin{equation}\label{E:32}
 m^2(q^2)=m^2_0+Re\Pi(q^2),\,Im\Pi(q^2)=q\Gamma(q^2)\,.
\end{equation}
In the case of unstable fermion we have another parametrization
scheme:
\begin{equation}\label{E:33}
 m(q^2)=m_0+Re\Sigma(q^2),\,Im\Sigma(q^2)=\Gamma(q^2)\,.
\end{equation}
So, the  definition of the  fermion function $\rho(\mu)$ demands
an additional analysis. If we choose for fermion UP the universal
density function (\ref{E:23}), which follows from convolution
method \cite{7}, then we must do exchange $Im\Pi(\mu)\rightarrow
\sqrt{\mu}Im\Sigma(\mu)$ in the Eq.(\ref{E:28}). Inserting the
result into Eq.(\ref{E:13}) with $\hat{D}(x,\mu)$, defined by
Eq.(\ref{E:12}), we can get the correspondence between the model
propagator of fermion unstable field and the effective theory one:
\begin{equation}\label{E:34}
 \int\frac{\hat{k}+\sqrt{\mu}}{k^2-\mu+i\epsilon}\,\rho(\mu)d\mu\longrightarrow
 \frac{\hat{k}+k}{k^2-m^2(k^2)-ik\Sigma(k^2)}\,,
\end{equation}
where $k=\sqrt{(kk)}$.  The numerator of the right side of
Eq.(\ref{E:34}) coincides with the expression $\hat{\eta}$ in Ref.
\cite{7}.

The transitions (\ref{E:24}), (\ref{E:31}) and (\ref{E:34})
establish the correspondence between the discussed model and some
effective theory of UP in the framework of traditional QFT
approach. These transitions follow from the determination of
$\rho(\mu)$, that is from the accounting of interaction, which
forms the wave packet (\ref{E:2}) and mass smearing. Above
mentioned effective theory has a close analogy with the
traditional description of UP in the intermediate state as a
special case of discussed approach. The most important feature of
the effective theory, chosen in such a way, is the possibility to
connect the decay-chain method and convolution method within the
framework of this theory \cite{7}. So, we have some
self-consistency of the discussed model, effective theory,
convolution and decay-chain method. However, due to some
difficulties, which arise in traditional approach, the search of
alternative $\rho(\mu)$ - definition is actual now.

\section{Phenomenological consequences of the model}

The phenomena of mass smearing take place on the various
hierarchical levels due to fundamental character of uncertainty
principle. The value of FWE in the particle physics depends on the
relations $\Gamma/M$ and $\Gamma/(M-M')$, where $M$ and $M'$ are
the masses of UP and total masses of decay products. So, FWE is
large in the decay or generation of UP with a large decay width
$\Gamma$ or in the near-threshold processes. There are many
examples of the particles with a large value of $\Gamma/M$ in the
hadron physics, for instance $\Gamma_{\rho}/M_{\rho}\approx 0.2$,
and we can observe a large effect in these cases. The fundamental
UP have, as a rule, negligible widths, except $Z,W$ bosons and $t$
quark, which have $\Gamma/M\sim 10^{-2}$. So, FWE can be
discovered by means of the precision measurements of decay
characteristics or in the near-threshold processes. In this
section we offer a short review of mass-smearing phenomena and
consider some examples of the processes, where FWE play a
significant role.

One of the most pure FWE in the hadron physics takes place in the
near-threshold decays $\phi(1020)\rightarrow K^{+}K^{-}, K^{0}_{L}
K^{0}_{S}$. The ratio of branchings does not depend on hadron
factors in a good approximation \cite{10} and is equal to the
ratio of phase space:
\begin{equation}\label{E:35}
 R=\frac{B(\phi\rightarrow K^{+}K^{-})}{B(\phi\rightarrow
 K^{0}_{L}K^{0}_{S})}=(\frac{1-4m^2_{+}/m^2_{\phi}}{1-4m^2_{0}/m^2_{\phi}})^{3/2},
\end{equation}
where $m_{+}=m(K^{\pm})$ and $m_{0}=m(K^{0})$. Inserting the
values of masses into Eq.(\ref{E:35}) we get the discrepancy
between theoretical and experimental $R$, which was discussed in
Ref. \cite{10}:
\begin{equation}\label{E:36}
 R^{th}=1.53 ;\,\,\,R^{exp}=1.45 \pm 0.03\,.
\end{equation}
The various corrections to $R^{th}$ have been calculated in
\cite{10} (Bramon et al), but the discrepancy remains (Fermi's
"Golden Rule" puzzle). Suggested model doesn't directly describe
FWE in the processes with fixed energy, for instance in the
process $e^+e^-\rightarrow\phi(1020)\rightarrow\bar{K}K$. Here we
consider model description of FWE in the process of type
$X\rightarrow Y\phi(1020)\rightarrow Y\bar{K}K$. The model
prediction with account of FWE gives the ratio in the form:
\begin{equation}\label{E:37}
 R^{M} =\frac{\int_{a_1}^{b}\Gamma^{+}_{\phi}(m)\rho(m)dm^2}
 {\int_{a_2}^{b}\Gamma^{0}_{\phi}(m)\rho(m)dm^2}\,,
 \end{equation}
where $a_1=4m^2_{+}$, $a_2=4m^2_{0}$, $b=E^2_{max}$,
$\Gamma^{a}_{\phi}(m)\sim m(1-4m^2_{a}/m^2)^{3/2}$,
$m_{a}=m(K^{a})$ and $a=0,\pm$. In the Breit-Wigner (BW)
approximation, which is applicable to narrow resonanses
($\Gamma_{\phi}/m_{\phi}\sim 10^{-3}$), the function $\rho(m)$ is
defined by the expression:
\begin{equation}\label{E:38}
 \rho(m)=\frac{1}{\pi}\,\frac{m_{\phi}\Gamma_{\phi}}{(m^2-m^2_{\phi})^2+m^2_{\phi}\Gamma^2_{\phi}}
\end{equation}
According to Eq.(\ref{E:37}) the value $R^{M}$ depends on upper
limit of integration $E_{max}$, which we have took in the interval
of two-particle generation:
\begin{equation}\label{E:39}
 R^{M}=1.43-1.41;\,\,\,\,\,\, E_{max}=(1.5-2.0)\, Gev.
\end{equation}
Thus, the model account of FWE in the process under consideration
gives the result, which is similar to experimental one
(\ref{E:36}). In the paper \cite{10}\,(Fischbach et al.) close
result was obtained with help of correction, caused by energy
dependence of matrix element. This approach has some analogy with
discussed treatment. With help of Eqs.(\ref{E:37}) and
(\ref{E:38}) one can evaluate the contribution of FWE into the
value $R$ for the decay channels $\phi(1020)\rightarrow
\rho\gamma$, $f_o(980)\rightarrow\bar{K}K$ and other.

Hadron decays of type $H\rightarrow H_1 H_2$ are direct objects of
suggested model, when $H_1$ and (or) $H_2$ are the hadrons with a
large width. The contribution of FWE into decay rates of the
decays $B^0\rightarrow D^-\rho^+$, $B^0\rightarrow D^-a^+_1$ and
$\Lambda^0_b\rightarrow \Lambda^+_c\rho^-$,
$\Lambda^0_b\rightarrow \Lambda^+_c a^-_1$ were evaluated \cite{6}
in the approach, which is similar to CM. The result of
calculations reveals that the contributions of FWE are large (from
20 to 40 percent) and its account improves the conformity of the
experimental data and theoretical predictions. Here we consider
the decay $B^0\rightarrow D^-\rho^+$ and evaluate FWE according to
discussed approach, which is in close analogy with one used in the
Ref.\cite{6} but gives other result. The two-body nonleptonic
decays of B mesons have been studied by Bauer et al. \cite{11} and
reanalysed with account of FWE in Ref.\cite{6}. Decay rate is
given by
\begin{equation}\label{E:40}
 \Gamma(B^0 \rightarrow
 D^-\rho^+)=\frac{\vert A(m_{\rho})\vert^2}{8\pi
 m^2_{\rho}}\,\,k^3\,\,,
\end{equation}
where $k$ is absolute value of final three-momentum in the rest
frame of the $B^0$ meson and $A(m_{\rho})$ is the decay amplitude:
\begin{equation}\label{E:41}
 A(m_{\rho})=\sqrt{2}G_Fa_1
 V_{cb}V^*_{ud}m_{\rho}F_{\rho}F_1(m^2_{\rho}).
\end{equation}
In the Eq.(\ref{E:41}) $a_1$ is Wilson coefficient, $f_{\rho}$ is
decay constant and $F_1(m^2_{\rho})$ is form factor at
$q^2=m^2_{\rho}$, which was approximated by a simple pole formula
$F_1(q^2)=F(0)/(1-q^2/m^2_{bc})$. The expression (\ref{E:41}) does
not include FWE and gives a marked difference between theoretical
and experimental values of branchings \cite{6}:
\begin{equation}\label{E:42}
 B^{th}=10.5\cdot 10^{-3}\,\,,\,\,\,\,B^{exp}=(7.5\pm 1.2)\cdot
 10^{-3}\,\,.
\end{equation}
The contribution of FWE into $B^{th}(B^0\rightarrow D^-\rho^+)$
was calculatrd in the Ref.\cite{6}:
\begin{equation}\label{E:43}
 \bar{B}^{th}=5.78\cdot 10^{-3},\,\,\,R=0.55.
\end{equation}
where $\bar{B}^{th}$ is branchings with account of FWE and
$R=\Bar{B}^{th}/B^{th}$. We recalculate the ratio $R$ taking into
consideration m-dependence of $f_{\rho}(m)$, $F_1(m)$ and
$\Gamma_{\rho}(m)$ (in analogy with approach \cite{6}). The
expression for $R$ has the form:
\begin{equation}\label{E:44}
 R=\int_{a}^{b}B^{th}(m)\rho(m)\,dm^2/B^{th}(m_{\rho}),
\end{equation}
where
\begin{align}
 \rho(m)=\frac{1}{\pi}\,\frac{m\Gamma_{\rho}(m)}{(m^2-m_{\rho}^2)^2+m^2\Gamma^2_{\rho}(m)},\\
 \Gamma_{\rho}(m)=(g_{\rho}^2/48\pi)m(1-4m_{\pi}^2/m^2)^{3/2},\\
 a=(2m_{\pi})^2,\,b=(m_B-m_D)^2.
\end{align}
The expressions $f_{\rho}(m)$ and $F_1 (m)$ are taken from
Ref.\cite{6}, the values $a$ and $b$ follow from the kinematics of
the $\rho$ and $B^0$ decays. With help of the Eqs.(44-47) we get
the result:
\begin{equation}\label{E:48}
 R=0.82,\,\,\,\bar{B}^{th}=8.64\cdot 10^{-3}.
\end{equation}
Thus, our approach leads to more realistic evaluation of FWE, that
is $R$, which improves the conformity of the experimental data and
theoretical predictions. There are many processes in the hadron
physics with participation of the hadrons with a large total
width, for instance $f_0(600)$, $\rho(770)$, $f_0(980)$,
$a_0(980)$ etc. In these cases we must take into account FWE,
particularly in the near-threshold processes. Another feature of
the mass-smearing phenomenon can manifests itself through the mass
dependence of the hadron factors, such as decay constant and form
factor, which have been taken into consideration in the discussed
decay $B^0\rightarrow D^-\rho^+ $.

The most of elementary (or fundamental) particles are unstable,
however, the large width have $W, Z$ bosons and $t$ quark only.
The ratio $\Gamma/M\sim 10^{-2}$ for these particles, that is the
value of FWE can be measured in the precision experiments or in
the near-threshold processes. The decay rates of the
near-threshold decays $t\rightarrow WZb,cWW$ and $cZZ$ were
calculated with account of FWE within the framework of CM and DCM
in the Refs.\cite{5}. The contributions of FWE lead to substantial
enhancement of decay rates, in particular of $B(t\rightarrow WZb)$
and $B(t\rightarrow cZZ)$. For instance, the branchings without
($B$) and with ($\bar{B}$) account of FWE in the first case differ
by an order of magnitude \cite{5} (Altarelli et al.):
$B(t\rightarrow WZb)\sim 10^{-7}$, $\bar{B}(t\rightarrow WZb)\sim
10^{-6}$. The description of these decays by suggested model does
not differ from the one in the Ref.\cite{5}. Here we consider the
contribution of FWE into decay rates of the decay channels
$W\rightarrow f_1\bar{f}_2$ and $Z\rightarrow f\bar{f}$. In the
approximation of massless fermion the partial width $\Gamma\sim
M^3$ and the ratio $R=\bar{B}/B$ is defined by the simple
expression:
\begin{equation}\label{E:49}
 R=\int_{a}^{b}m^3 \rho(m)\,dm^2/M^3.
\end{equation}
In the case of process $e^+e^-\rightarrow Z\rightarrow W^+W^-$
near the threshold ($\sqrt{s}\approx 2M_W$) $a\approx m^2_f$,
$b\approx s$. In the Breit-Wigner approximation for $\rho(m)$ from
the Eq.(\ref{E:49}) we get $R_W\approx 1.04$. The same result
takes place for the $Z$-pair generation near threshold. It should
be noted, that the values of the limits of integration in the
Eq.(\ref{E:49}) crucially depends on the process of $W$ or $Z$
generation. In the processes of type $e^+e^-\rightarrow
Z\rightarrow f\bar{f}$ mass parameter is fixed $m=\sqrt{s}$ and we
have the decay properties of $Z$ as function of $\sqrt{s}$. Thus,
the contributions of FWE into decay properties of $Z$, $W$ bosons
and $t$ quark must be taken into consideration in the precision
measurements. The evaluation of these contributions can be
fulfilled in the framework of the convolution method in a simple
way. It should be noted that considered effects don't influence on
the precision measurements of $Z$ properties at fixed energy
$\sqrt{s} \approx M_Z$.

The effect of the mass smearing can plays a significant role in
the mixing, oscillation and CP violation in the systems of neutral
mesons $M^0-\bar{M}^0$. Large contribution of FWE into mixing is
due to the width of the short-lived state is comparable with the
splitting of mass $\Gamma_S\sim\Delta m=\vert m_S-m_L\vert $. So,
the levels of the short-lived ($M_S$) and long-lived ($M_L$)
states strongly overlap due to mass smearing and this effect can
influence on the mixing. To illustrate this phenomenon we compare
the values $\Gamma$ and $\Delta m$ in the case of $K^0-\bar{K}^0$
and $B^0-\bar{B}^0$ systems \cite{11}:
\begin{align}
 \Gamma(K^0_S)&=7.30\cdot 10^{-6} eV,\,\,\,\Delta m_K=3.48\cdot 10^{-6} eV, \notag \\
 \Gamma(B^0_d)&=4.24\cdot 10^{-4} eV,\,\,\,\Delta m_{B_d}=3.34\cdot 10^{-4}
 eV, \\
 \Gamma(B^0_s)&=4.43\cdot 10^{-4} eV,\,\,\,\Delta m_{B_s}\geq 94.8\cdot10^{-4}
 eV \notag
\end{align}
In the case of $B^0-\bar{B}^0$ systems short- and long-lived
components have near the same widths and the division is usually
marked by $M_H$ (heavy) and $M_L$ (light) states. Now we consider
the phenomenological consequences of mass smearing in the systems
of neutral mesons.

The theoretical evaluations of mass splitting, which are based on
short distance FCNC transition (box diagrams), do not account FWE.
So, the contradictions can take place between the theoretical
$\Delta m^{th}$ and experimental values $\Delta m^{exp}$, which
follow from the oscillation experiments. Here we demonstrate this
on the examples of $K^0-\bar{K}^0$ and $B^0-\bar{B}^0$ systems.

The experimental value of mass splitting $\Delta m^{exp}$ follows
from the observable characteristic of oscillation $\chi$:
\begin{equation}\label{E:51}
 \chi=\frac{x^2}{2(1+x^2)},\,\,\,x=\frac{\Delta
 m^{exp}}{\Gamma}\,\,\,(when \,\,\,y=\frac{\Delta \Gamma}{\Gamma}\ll x).
\end{equation}
When $\Gamma_S\sim \Delta m^{exp}$, then the measured $\chi$ has
some effective value, averaged by the mass distribution:
\begin{equation}\label{E:52}
 \bar{\chi}=\int \chi(m)\rho(m)\,dm^2\,,
\end{equation}
where according to Eq.(\ref{E:51}) $\chi(m)=\Delta m^2(m)/2(\Delta
m^2(m)+\Gamma^2)$ and $\Delta m(m)=\vert m-M_L\vert$. The value
$\bar{\chi}$ can substantially differs from $\chi_0=\Delta
m^2_0/2(\Delta m^2_0+\Gamma^2)$, where $\Delta m_0=\vert
m_L-\bar{m}_S\vert $ and $m_L$ is constant. As a result the mass
splitting, which follows from the oscillation experiment, can
significantly differs from the theoretical value $\Delta m_0$,
which follows from the FCNC transitions. Here we consider this
effect for the $K^0-\bar{K}^0$ system in more detail. The
theoretical value $\Delta m^{th}_K$ is defined by the expression
\cite{12}:
\begin{align}\label{E:53}
 \Delta m^{th}_K &= \frac{G^2_f}{6\pi^2}\,m^2_W m_K B_K F^2_K
          ( \eta_1\vert
                 U^*_{cs}U_{cd}\vert ^2s_0(x_c)\\
                 &+ \eta_2 \vert U^*_{ts}U_{td}\vert ^2 s_0(x_t)+2\eta_3\vert
                 U^*_{cs}U_{cd}U^*_{ts}U_{td}\vert s_0(x_c,x_t)
          ). \notag
\end{align}
In the Eq.(\ref{E:53}):
\begin{align}\label{E:54}
 & s_0(x_c)= x_c,\,\,\,s_0(x_t)=\frac{4x_t-11x^2_t+x^3_t}{4(1-x_t)^2}
 -\frac{3x^2_t\ln{x_t}}{2(1-x_t)^3},\,\,\, x_c=\frac{m^2_c}{m^2_W},\,\,\,\\
 &
 x_t=\frac{m^2_t}{m^2_W},\,\,\,s_0(x_s,x_t)=x_c(\ln{\frac{x_t}{x_c}}-\frac{3x_t}{4(1-x_t)}-\frac
 {3x^2_t\ln{x_t}}{4(1-x_t)^2}). \notag
\end{align}
The most detailed evaluation of hadron factors, which enter to
Eq.(\ref{E:53}), was fulfilled in Ref.\cite{12}, including the
corrections beyond leading logarithm:
\begin{equation}\label{E:55}
 B_K=0.84,\,\eta_1=1.38,\,\eta_2=0.574,\,\eta_3=0.47,\,F_K=160
 MeV.
\end{equation}
Using (\ref{E:55}) and other data from \cite{11} with help of
(\ref{E:53}) we get $\Delta m^{th}_K$, which is substantially less
then $\Delta m^{exp}_K$:
\begin{equation}\label{E:56}
 \Delta m^{th}_K=2.34\cdot 10^{-6} eV,\,\,\,\Delta
 m^{exp}_K=3.48\cdot 10^{-6} eV.
\end{equation}
The characteristics of oscillation, which correspond to these mass
splitting, are defined by Eq.(\ref{E:51}):
\begin{equation}\label{E:57}
 \chi^{th}_K=0.047,\,\,\,\chi^{exp}_K=0.093
\end{equation}
This large discrepancy can be eliminated by accounting of the
mass-smearing effect. In the BW approximation Eq.(\ref{E:52}) can
be rewritten in the form:
\begin{equation}\label{E:58}
 \bar{\chi}=\frac{\Gamma}{\pi}\int_{a}^{b}\frac{(x+\Delta
 m)^2\,dx}{((x+\Delta m)^2+\Gamma^2)(4x^2+\Gamma^2)}\,,
\end{equation}
where $x=m-m_S$, $\Gamma=\Gamma_S$ and the limits of integration
$(a,b)$ are defined by the interval of $m$ variation, which,
however, is not correctly determined. To evaluate $\bar{\chi}$ we
fix the interval $m_L\leq m \leq m_S+n\Delta^{th} m$, where
$n=5,...,100$. Inserting $\Delta m=\Delta^{th}m$ into
Eq.(\ref{E:58}) we get:
\begin{equation}\label{E:59}
 \bar{\chi}_K=0.075-0.117
\end{equation}
Thus, the account of mass smearing in the $K^0-\bar{K}^0$ mixing
can significantly change the value $\chi$ and make it possible to
establish the accordance between $\chi^{exp}_K=0.093$ and
$\Delta^{th}m_K=2.34\cdot 10^{-6} eV$. The value of the mass
splitting in the $B^0_d-\bar{B}^0_d$ system follows from the
expression \cite{12}:
\begin{equation}\label{E:60}
 \Delta m_B =\frac{G^2_F}{6\pi^2}\, m^2_WM_B\eta _B B_B f^2_B
 s_0(m^2_t/m^2_W)(V_{td} V ^*_{tb})^2,
\end{equation}
where:
\begin{equation}\label{E:61}
 s_0(x_t)=0.784x^{0.76}_t,\,\,\,f_B\sqrt{B_B}=220
 MeV,\,\,\,\eta_B=0.551
\end{equation}
Using the rest data from \cite{11}, we get the central value of
$\Delta m^{th}_B$, which is slightly larger the experimental one:
\begin{equation}\label{E:62}
  \Delta m^{th}_B=3.51\cdot 10^{-4} eV,\,\,\,\Delta
 m^{exp}_B=3.34\cdot 10^{-4} eV.
\end{equation}
Inserting the value $\Delta m^{th}_B$ into Eq.(\ref{E:58}) we get
for the same interval of $m$ variation:
\begin{equation}\label{E:63}
 \bar{\chi}^{th}_B=0.18-0.20,\,\,\,\chi^{exp}_B=0.188\pm 0.003
\end{equation}
Thus, we get the realistic result, but for the correct evaluation
of FWE contribution in this case we need more precise measurement
of CKM element $V_{td}$. The effect of mass smearing in
$B^0_s-\bar{B}^0_s$ system is small due to inequality $\Delta m_B
\gg \Gamma _B$. In this case from Eq.(\ref{E:51}) we have
$\chi\approx 0.5\approx\bar{\chi}$. It should be noted, that
investigation of the mass-smearing effect (or FWE) in the
$M^0-\bar{M}^0$ systems demands more detailed and self-consistent
analysis, which must include the fit of CKM mixing matrix. The
above considered examples are the illustrations, which can
stimulate further investigation of the discussed effect. The same
conclusion takes place for the indirect CP violation, which caused
by the mixing.

In this section we have considered some examples of processes in
the various fields of particle physics, were FWE give large
contributions. We fulfilled rough evaluations of these
contributions to illustrate the important role of FWE in some
specific cases. The conclusion follows from this short analysis,
that mass-smearing effect should be taken into consideration in
the hadron and particle physics.

\section{Conclusions}

The finite-width effects in the processes with participation of UP
can be described by renormalized propagator, decay-chain method,
convolution method and effective theory of UP. The convolution
formula is convenient instrument for calculations of decay rate
and gives the results in accordance with experiment. In this paper
we have considered the model of UP with a random mass and derived
the convolution formula as a direct consequence of the model. The
model operator function and Lagrangian have a convolution
structure and describe the mass-smearing effects in accordance
with the uncertainty principle.

The principal element of suggested model is probability density
function $\rho(\mu)$, which describes the main properties of UP.
Traditional description of UP in the intermediate state by
resonance line with complex pole (or by dressed propagator with
mass and width as parameters) corresponds to the model description
of UP in arbitrary state by function $\rho(\mu)$ with the same
parameters. We have considered the determination of $\rho(\mu)$
from DCM and by matching the model propagator to renormalized one.
The first approach is equivalent to the convolution method or
truncated decay-chain method. The second one has some
restrictions, caused by propagator renormalization peculiarities.
It should be noted, that the mass-smearing effect follows from the
fundamental uncertainty principle, then the search of $\rho(\mu)$
from the first principles is reasonable.

We have considered some examples of FWE manifestations in a
various regions of the particle physics. The fulfilled analysis
and evaluations of the FWE contributions to observable
characteristics lead to the conclusion: these contributions can be
large, an account of it improves the conformation of experimental
data and theoretical predictions, convolution method gives a
simple and convenient tool for evaluation of the effect.

\end{document}